\documentclass[runningheads]{llncs}
\usepackage{graphicx}
\usepackage{epstopdf}
\usepackage{multirow}
\usepackage{algorithm}
\usepackage[compatible]{algpseudocode}
\usepackage[T1,T2A]{fontenc}
\usepackage[utf8]{inputenc}
\graphicspath{{pictures/}}
\DeclareGraphicsExtensions{.pdf,.png,.jpg,.eps}

\begin{document}
\title{Optimal placement of mobile distance-limited devices for line routing\thanks{The study was carried out within the framework of the state contract of the Sobolev Institute of Mathematics (project FWNF-2022-0019)}}
\author{Adil Erzin\inst{1,2,3}\orcidID{0000-0002-2183-523X}\and Anzhela Shadrina\inst{2}}
\authorrunning{A. Erzin, A. Shadrina}
\institute{Sobolev Institute of Mathematics, SB RAS, Novosibirsk 630090, Russia \and
Novosibirsk State University, Novosibirsk 630090, Russia\\
\email adilerzin@math.nsc.ru}
\titlerunning{Optimal placement of mobile distance-limited devices...}
\maketitle              
\begin{abstract}
 A segment (barrier) is specified on the plane, as well as depots, where the mobile devices (drones) can be placed. Each drone departs from its depot to the barrier, moves along the barrier and returns to its depot, traveling a path of a limited length. The part of the barrier along which the drone moved is \emph{covered} by this sensor. It is required to place a limited quantity of drones in the depots and determine the trajectory of each drone in such a way that the barrier is covered, and the total length of the paths traveled by the drones is minimal.

 Previously, this problem was considered for an unlimited number of drones. If each drone covers a segment of length at least 1, then the time complexity of the proposed algorithm was $O(mL^3)$, where $m$ is the number of depots and $L$ is the length of the barrier. In this paper, we generalize the problem by introducing an upper bound $n$ on the number of drones, and propose a new algorithm with time complexity equals $O(mnL^2)$. Since each drone covers a segment of length at least 1, then $n\leq L$ and $O(mnL^2)\leq O(mL^3)$. Assuming an unlimited number of drones, as investigated in our prior work, we present an $O(mL^2)$-time algorithm, achieving an $L$-fold reduction compared to previous methods. Here, the algorithm has a time complexity that equals $O(L^2)$, and the most time-consuming is preprocessing.
\end{abstract}
\section{Introduction}
Monitoring of extended objects (roads, pipelines, power lines, etc.) must be performed regularly to check their condition. It is not always possible and/or advisable for a human to do this. Mobile autonomous devices, which we will call the \emph{drones}, are used for monitoring, especially in hard-to-reach areas. A drone is equipped with the appliance for monitoring. In the vast majority of publications related to the topic, a drone or sensor collects data within a certain area, often a disk. Here, it is assumed that the drone has \emph{covered} this area. If the object to be covered is extended, then effective monitoring is called a barrier coverage problem, where a barrier is understood as a line segment. Drones, naturally, have a limited energy supply, which entails a limitation on the maximum length of the path traveled without renewing the energy. The literature considers the problems of minimizing the number of drones in covering a barrier (MinNum), minimizing the maximum length of the path traveled by each drone (MinMax), and minimizing the total length of paths traveled by drones (MinSum).

In this paper, we consider a MinSum problem, since the MinNum and MinMax problems are solved in \cite{ErShad24}. The problem MinSum receives dedication from various publications \cite{Andrews:17,Bar:15,Benk:15,Cherry:17,Czyzowicz:10,Er:18,Err:18,Wang:15}. In the MinSum problem considered in the overwhelming majority of publications, each drone has the initial location and radius of the disk that it covers. It is required to find a subset of drones and their final positions such that the barrier is covered, and the total length of the paths traveled by drones is minimal \cite{Bhattacharya:09,Cherry:17,Czyzowicz:10,Dobrev:15,Er:18,ErLagIr:19}. The MinSum problem is NP-hard with different disks \cite{Czyzowicz:10,Dobrev:15}. If the disks involved in the coverage are equal, then the complexity status of the problem is not known. However, in two-dimensional (2D) Euclidian space an $O(n^4)$-time algorithm is known, where $n$ is the number of sensors, that builds a $\sqrt{2}$-approximate solution \cite{Cherry:17}, the complexity of which was reduced to $O(n^2)$ in \cite{Er:18}.

In this paper, we consider a 2D MinSum problem in a slightly different formulation. A barrier as a line segment and a set of arbitrarily arranged depots are on the plane. Each drone covers only the point at which it is located. Then, in order to cover a part of the barrier, the drone must move along this part. After that, the drone must return to its depot. Here, the length of the path traveled by the drone must not exceed the specified value, which is determined by its energy reserve.

Over the past years, drones have become more and more popular. However, the complexity of routing drones has not been fully investigated. Coutinho et al. defined the Routing and Trajectory Optimization Problem for drones in their 2018 publication \cite{Coutinho:18}. They introduce a taxonomy and review recent contributions in drone trajectory optimization, drone routing, and papers addressing these problems. In \cite{Campbell:18}, the authors present the Drone Arc Routing Problem and study its relation with well-known Postman Arc Routing Problems. The paper \cite{Kek:08} proposes two distance-constrained vehicle routing problems, and study potential benefits in flexibly assigning start and end depots. Drones provide reconnaissance support for the military and often need operational routes immediately. Current practice involves manual route calculation that can involve hundreds of targets and a complex set of operational restrictions. The research in \cite{Kinney:17} focused on providing an operational drone routing system.

In \cite{ErPlot20}, some properties of the solution for the problem under consideration, when in each depot contains only one drone, were presented. The authors proposed an implementation of the dynamic programming algorithm for constructing an optimal cover in the Manhattan metric and also find special cases when this coverage is the optimal solution to the MinNum problem.

\subsection{Our Contribution}
Unlike \cite{ErPlot20}, in this paper, as well as in \cite{ErShad24}, the number of drones in each depot is the desired value. In \cite{ErShad24}, the number of drones in each depot was unlimited, and we proposed the algorithms for solving the MinNum, MinMax, and MinSum problems. Under the assumption that each drone covers a segment of length at least 1, for the MinSum problem in \cite{ErShad24}, an $O(mL^3)$-time algorithm was developed, where $m$ is the number of depots and $L$ is the barrier length. The most time-consuming step was preprocessing, during which the minimum length of drone paths from each depot covering the segment $[a,b]$ for all $a,b\in [0,L]$ was calculated. In this paper, we consider a generalization of the MinSum problem when the total number of drones is limited to $n$. We propose two versions of the dynamic programming algorithm. Without preprocessing, the first algorithm finds the optimal coverage by the \emph{optimal} number of drones $n^*$ with time complexity equal to $O(L^2)$. If $n^*\leq n$, then this solution is optimal for the case of the upper bounded number of drones as well. Otherwise, the second $O(mnL^2)$-time two-parameter dynamic programming algorithm is used to construct the optimal cover. It was possible to reduce the complexity by more efficient preprocessing, in such a way that the complexity became equal to $O(mnL)$. Since each drone covers the segment of length at least 1, then $n\leq L$. As a result, we reduced the complexity of solving the problem by $L$ times when the number of drones is not limited, and kept the same complexity for more general problem than the problem considered in \cite{ErShad24}.\\

The rest of this paper is organized as follows. Section 2 provides the formulation of the problems. In Section 3, algorithms for building the solutions are proposed. In Section 4, we propose a new method for calculating the minimum path length of drones at one depot covering different segments of the barrier more efficiently than the algorithm described in \cite{ErShad24}. Section 5 concludes the paper.

\section{Problem Formulation}
On the plane on the $0X$ axis there is a barrier -- the segment $[0,L]$. The coordinates of all depots $d_i=(x_i,y_i)$, $y_i\geq 0$, $i\in D$, $|D|=m$, in which at most $n$ drones can be placed, are given. We assume that $x_i\neq x_j$ if $i\neq j$, since otherwise, the depot with a greater ordinate can be excluded. We number the depots from left to right according to their abscissas, i.e. $i<j$ if and only if $x_i<x_j$. Let $q>0$ be the maximum path length that any drone can travel. Each drone departs from its depot, moves along the barrier and returns to its depot. We say that the drones from depot $i$ \emph{cover} the segment $[a_i,b_i]\subseteq [0,L]$ if each of them moves from the depot $d_i$ to the barrier point, travel along the barrier and return to its depot, following a path whose length is at most $q$, and collectively cover the segment $[a_i,b_i]$.

\begin{definition}
The cover  $C$ of the barrier $[0,L]$ is the set of such segments $[a_i,b_i]\subseteq[0,L]$, that each of them is covered by drones of one depot $i\in D$, $\cup_{i\in D}[a_i,b_i]\supseteq [0,L]$, and the path length of each drone is upper bounded by $q$.
\end{definition}

To specify the cover $C$, it is necessary to determine the number of drones in each depot, the segments $[a_i,b_i]$ that are covered by the drones of the depot $i\in D$, as well as a trajectory of each drone.

\textbf{\underline{Problem MinSum.} It is required to place at most $n$ drones in $m$ depots in such a way as to cover the entire barrier $[0,L]$ and so that the total length of the drones’ paths is minimal.}

In the Euclidean metric, the optimal trajectory of any drone is a triangle with one vertex in the depot, with one side on the barrier and whose perimeter does not exceed $q$. Evidently, in the optimal cover the barrier segments covered by different drones do not overlap.

\section{Algorithms}
This section provides the algorithms for solving the problem under consideration. First, we must check the existence of the solution. For each depot $i$ we can determine a max-length segment $[A_i,B_i]$ in the barrier that can be covered by drones of the depot $i$. We assume that each drone covers a segment of length at least 1, then we can use the equations (1) and (2) from \cite{ErShad24} to calculate $A_i$ and $B_i$. Obviously, for the existence of the cover, the condition $\cup_{i\in D} [A_i,B_i]\supseteq [0,L]$ must be satisfied. In addition, the number of drones covering the barrier must not exceed $n$. To check this property, we can run the algorithm $A_{MinNum}$ from \cite{ErShad24}, which determines the minimum number $n_{min}$ of drones covering the barrier. The solution exists if $n_{min}\leq n$.

We will need the following definitions and statements.

\begin{lemma}\cite{ErShad24}
 There is an optimal cover in which drones from one depot cover one connected segment.
\end{lemma}

\begin{definition}\cite{ErShad24}
 A cover in which for all $i,j\in D$ we have $b_i\leq a_j$ if and only if $i<j$ is called an \emph{order-preserving cover} (OPC).
\end{definition}

\begin{lemma}\cite{ErShad24}
 There is always an optimal OPC for the MinSum problem.
\end{lemma}

Let $g(k,a,b)$ be the minimum total length of paths of $k$ drones in one depot that cover the segment $[a,b]$.

\begin{lemma}\cite{ErShad24}
 For any fixed segment $[a,b]$, the function $g(k,a,b)$ increases monotonically with an increase of $k$.
\end{lemma}

From the last lemma, follows that in the optimal cover, any segment covered by drones from one depot will be covered by the minimum number of drones. Determining this minimum number of drones in the deport $i$, let us define it as $n_i(a,b)$, is simple, assuming that all of them, except perhaps one, travel the maximum distance $q$.

In \cite{ErShad24}, functions $f_i(a,b)\geq 0$ were introduced that are equal to the minimum length of $n_i(a,b)$ drones paths from depot $i$ that cover the segment $[a,b]$ ($f_i(a,a)=0$). Suppose that we know the values of functions $f_i(a,b)$ and $n_i(a,b)$ for all $i=1,\ldots,m$ and integers $a=0,1,\ldots,L-1$ and $b=a+1,\ldots,L$. Let us introduce a new function equal to the minimum length of drone paths covering the segment $[a,b]$
$$
 f(a,b)=\min\limits_{i\in D} f_i(a,b)=f_{k(a,b)}(a,b).
$$
Here, there is a depot $k(a,b)\in D$, $n_{k(a,b)}(a,b)$ drones from which cover the segment $[a,b]$ traveling together the min-length distance. Thus, any partition of the barrier into the segments determines the cover. Therefore, it is necessary to solve the following problem of optimal partition of the segment (interval).

\begin{equation}\label{e1}
  \sum_{i=1}^{p} f(z_{i-1},z_i)\rightarrow\min\limits_{z,p};
\end{equation}
\begin{equation}\label{e2}
  0=z_0<z_1<\ldots,z_p=L,
\end{equation}
where unknown $p$ is the number of subsegments, i.e. here, it is required to find the optimal partition of the barrier into the \emph{optimal} number of parts.

The problem (\ref{e1})-(\ref{e2}) is well studied. A dynamic programming algorithm was proposed to solve it no later than in the 1970s (see, for example, the monograph \cite{Beresnev:78}). Later, the algorithm has been rediscovered several times \cite{Jackson05,Tilli:19,Vidal:93}. To describe the algorithm, let us call it $A_1$, we assume that $L$ and parameter $l\in [0,L]$ are integers. Let $S(l)$ be the minimum total paths length of drones covering the segment $[0,l]$, $l=0,1,\ldots,L$.

\subsection{Algorithm $A_1$}
This is a dynamic programming algorithm based on the Bellman principle that in an optimal partition of a segment, any subsegment is also optimally partitioned. The algorithm runs the forward and backward recursions. During the forward recursion, for each value $l=0,1,\ldots,L$, the rightmost (last) partition point $z(l)<l$ is found such that the total length of the drone’s paths covering the segment $[0,l]$ is minimal.

\textbf{Forward recursion.}
\begin{equation}\label{e3}
  S(0)=0;\ S(l)=\min\limits_{0\leq z< l} \{f(z,l)+S(z)\}=f(z(l),l)+S(z(l)),\ l=1,\ldots,L.
\end{equation}

We keep in memory the rightmost (conditionally optimal) point of the partition $z(l)<l$ for all $l=1,\ldots,L$.

\textbf{Backward recursion.} At this stage, we will find the number of subsegments $p^*$ in the optimal partition, as well as the optimal partition itself. Set the optimal number of drones $n^*=0$ and the number of subsegments $p^*=0$. At the end of forward recursion, when $l=L$, we found the minimal value $S^*=S(L)$ of the objective function (\ref{e1}) and the last optimal partition point $z(L)$. Then segment $[z(L),L]$ is covered by $n_{k(z(L),L)}(z(L),L)$ drones from the depot $k(z(L),L)$, and we set $p^*=p^*+1$ and $n^*=n^*+n_{k(z(L),L)}(z(L),L)$. We set the new right endpoint of the segment as $l=z(L)$. Since during forward recursion we remember $z(l)$ for all $l=1,\ldots,L$, we know the last partition point of the segment $[0,l]$, which is the penultimate point of the optimal partition of the segment $[0,L]$. If $z(l)>0$, then we set $p^*=p^*+1$, $n^*=n^*+n_{k(z(l),l)}(z(l),l)$, $l=z(L)$ and continue finding the optimal partition points until $z(l)=0$. As a result, the optimal partition $z^*$ of the barrier into the $p^*$ parts and optimal number of drones $n^*$ will be found.

Obviously, formula (\ref{e3}) can be used if the variables are real. However, in this case, we will have a continuum of function values. If the variables are integers, then the complexity of the algorithm is a linear function regarding the input length $O(L^2)$, which is determined by the number of values of the function $f(a,b)$. The complexity of this algorithm is $m$ times less than the complexity of the algorithm $A_{MinSum}$ proposed in \cite{ErShad24} for solving the similar problem. Here we say nothing yet about the complexity of finding the values of the functions $f_j(a,b)$. A less time-consuming method for calculating them than in \cite{ErShad24} will be described below in the corresponding section.

If $n^*\leq n$, then the MinSum problem is solved. Otherwise, we apply the following algorithm $A_2$.

\subsection{Algorithm $A_2$}
The algorithm described below is applied if the optimal number of drones found by the algorithm $A_1$ is $n^*>n$. This is an infeasible solution. To construct a feasible solution, we need the following implementation of two-parameter dynamic programming. Let function $S_i(l,N)$ be equal to the minimum length of paths of $N$ drones from the depots $1,\ldots,i$ covering segment $[0,l]$. It follows from Lemma 3 that if a segment can be covered by drones from one depot, then it is covered by the minimum number of drones. Therefore, if the partition of the barrier is known, then the number of drones involved in covering the barrier is found uniquely. Obviously, the number of drones involved in the optimal coverage may be less than $n$. Let $N_i(l,N)\leq N$ be the number of drones in the first $i$ depots, where the parameter $N=0,1,\ldots,n$ is the maximum number of drones that can be placed in the depots $1,\ldots,i$.

In the solution constructed by the $A_2$ algorithm, the barrier is divided into $m$ segments. Here, some partition points $0=z_0\leq z_1\leq\ldots\leq z_m=L$ may coincide, i.e. some depots will not accommodate any drones. This means that this depot covers an empty segment.

\textbf{Forward recursion.}
For any $i=1,\leq,m$, $l=0,1,\ldots,L$ and $N=0,1,\ldots,n$, we calculate

\begin{equation}\label{e4}
 S_1(l,N)=
 \left\{
  \begin{array}{ll}
    f_1(0,l), & \hbox{if $N_1(l,N)=n_1(0,l)\leq N$;} \\
    +\infty, & \hbox{otherwise.}
  \end{array}
 \right.
\end{equation}

\begin{equation}\label{e5}
 \begin{array}{ll}
    S_i(l,N)=\min\limits_{z=0,\ldots,l}\{f_i(z,l)+S_{i-1}(z,N-n_i(z,l))\}= \\
    f_i(z_i(l,N),l)+S_{i-1}(z_i(l,N),N-n_i(z_i(l,N),l))
  \end{array}
\end{equation}
if $N_i(l,N)=n_i(z_i(l,N),l)+N_{i-1}(z_i(l,N),N-n_i(z_i(l,N),l))\leq N$ and $S_i(l,N)=+\infty$, otherwise.

Let us store in memory $z_i(l,N)$, $i=1,\ldots,m$, $l=0,1,\ldots,L$ and $N=1,\ldots,n$.

\textbf{Backward recursion.}
At the end of the forward recursion, with $N=n$, $l=L$ and $i=m$, the optimal value of the functional $S^*=S_m(L,n)$, the number of drones $N^*=N_m(L,n)$, and the last (rightmost) partition point $z_{m-1}^*=z(l)$ are found. Then, the segment $[z_{m-1}^*,L]$ (which may be empty) is covered by $n_m(z_{m-1}^*,L)$ drones of depot $m$. This means that the segment $[0,z_{m-1}^*]$ is covered by drones from depots $1,\ldots,m-1$. Let $l=z_{m-1}^*$. During the forward recursion, we found and remembered $z_i(l,N)$ for all parameter values, so the next optimal partition point from the right is $z_{m-2}^*=z(l)$. Continuing this process, we will find the optimal partition of the original barrier into the $m$ parts.

Naturally, recurrence relations (\ref{e4})-(\ref{e5}) are valid for any values of the parameters. However, for a polynomial implementation of the algorithm, we require that the parameters be integers. Then the time complexity of the algorithm $A_2$ is equal to $O(mnL^2)$.

\section{Calculating the functions $n_i(a,b)$ and $f_i(a,b)$}
To implement the algorithms described above, it is necessary to know the functions $f_i(a,b)$ and $n_i(a,b)$ for all $i=1,\ldots,m$, $a=0,1,\ldots,L-1$ and $b=a+1,\ldots,L$, i.e. $O(mL^2)$ values. Finding these values is a preprocessing before applying algorithms $A_1$ and $A_2$, which mostly determines the complexity of solving the problem. In \cite{ErShad24}, a method that finds $f_i(a,b)$ with complexity $O(mnL^2)$ was proposed. Let us show that calculating the values of the functions $f_i(a,b)$ and $n_i(a,b)$ for each depot $i\in D$ can be done with time complexity $O(nL)$. First, we determine $n_i(a,b)$. Since $O(n)$ drones can cover the segment, then if for each pair of values $a$ and $b$ we find $n_i(a,b)$, then $O(nL^2)$ operations are required. To reduce the complexity, we will calculate $n_i(a,b)$ differently. For each fixed value $a\in\{0,1,\ldots,L-1\}$ we will find a segment of \emph{maximum} length $[a,b(n_i)]$, which is covered by $n_i\in\{1,\ldots,n\}$ drones. Since the optimal solution for covering the segment $[a,b]$ uses the minimum number of drones from one depot, we can determine $n_i(a,b)$ for any $b$. To do this, we set $n_i(a,b)=n_i$, $b\in(b(n_i-1),b(n_i)]$. First, we find the segment $[a,b(1)]$, which covers one drone traveling the path of maximum length $q$. The covered segment includes some number of integers $b\in(a,b(1)]$. Knowing $b(1)$, we find the segment $[b(1),b(2)]$ of the maximum length, which covers the second drone. Continuing adding drones, we determine the minimum number of drones covering the segment $[a,b]$ for all integers $b=a+1,\ldots,L$. Thus, to calculate $n_i(a,b)$ for all $i=1,\ldots,m$, $a=0,\ldots,L-1$ and $b=1,\ldots,L$, $O(Lnm)$ operations are required.

\begin{figure}
\centering
\includegraphics[bb= 0 0 800 160,clip, scale=0.5]{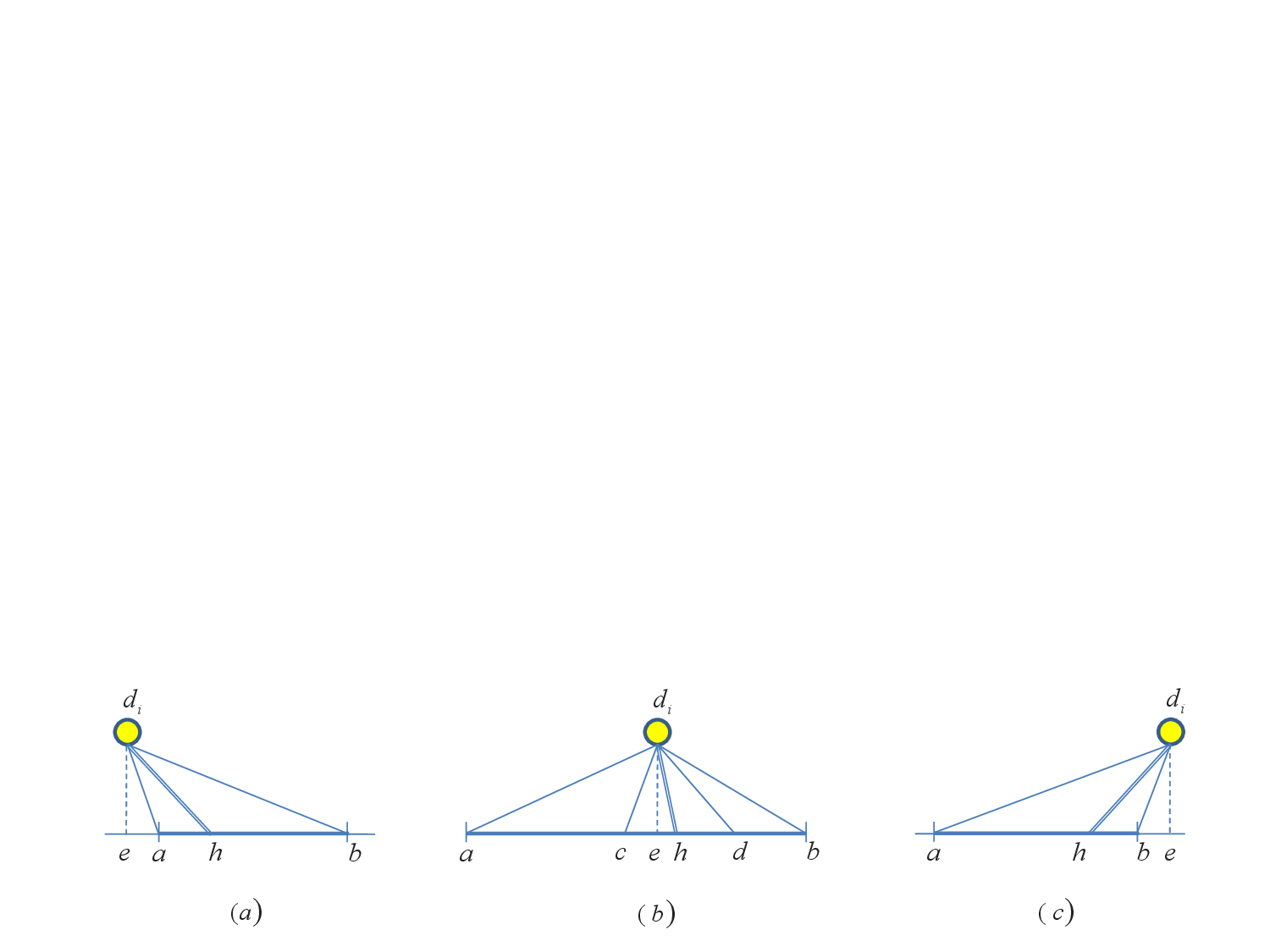}
\caption{Illustration for calculating the $f_i(a,b)$}
\label{fig1}
\end{figure}

Knowing $n_i(a,b)$, we calculate $f_i(a,b)$. For each depot $p=(x,y)$ three cases are possible: (\emph{a}) $x<a$, (\emph{b}) $x\in[a,b]$ and (\emph{c}) $x>b$ (Fig. \ref{fig1}). In cases (\emph{a}) and (\emph{c}), each of the $n_i(a,b)-1$ drones travels a maximal distance of length $q$, covering together the segment $[h,b]$ in case (\emph{a}) and the segment $[a,h]$ in case (\emph{c}), and one drone covers the segment $[a,h]$ traveling along the edges of the triangle $aph$ in case (\emph{a}) (Fig. \ref{fig1}\emph{a}) and the segment $[h,b]$ moving along the edges of the triangle $hpb$ in case (\emph{c}) (Fig. \ref{fig1}\emph{c}). In case (\emph{b}), each of the $n_i(a,b)-1$ drones travels a path of length $q$. It remains to determine the minimum length of the drone paths covering the segment $[c,d]$ (Fig. \ref{fig1}\emph{b}). Here, the segments $[a,c]$ and $[d,b]$ are covered by $n_i(a,b)-2$ drones, each of which passes a distance equal to $q$. We found $n_i(a,c)$ and $n_i(d,b)$ earlier. If the segment $[c,d]$ can be covered by one drone, then we cover it with one drone. Otherwise, it will be covered by two drones so that the line $[h,q]$ has a minimum slope (i.e. it is as close as possible to the perpendicular $pe$. The point $h$ is easy to find. If one drone can cover both segments $[c,e]$ and $[e,d]$, then $[p,h]$ is the perpendicular, i.e. $h=e$. Otherwise, one of these segments cannot be covered by one drone. Let this be the segment $[e,d]$. Then one drone covers the segment of maximum length $[h,d]$, and the other one covers the remaining segment $[c,h]$.

The time complexity of computing $n_i(a,b)$ and $f_i(a,b)$ for all $i\in D$, $a$ and $b$ is $O(Lnm)$. If the number of drones is unknown, then by the assumption that each drone covers a segment of length at least 1, the complexity is $O(mL^2)$.

\subsection{Illustration}
\begin{figure}
\centering
\includegraphics[bb= 0 0 550 300,clip, scale=0.5]{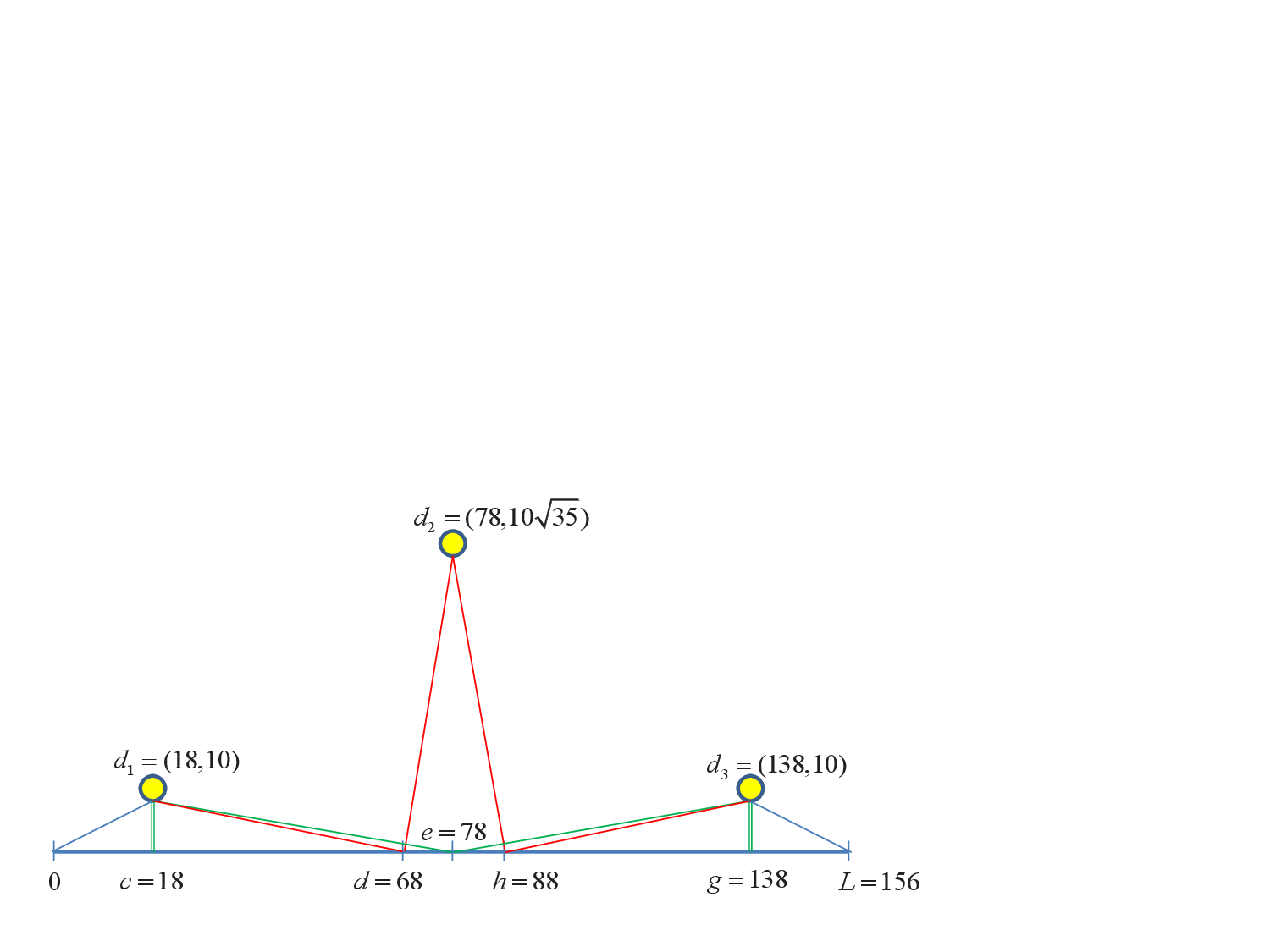}
\caption{Illustration for example}
\label{fig2}
\end{figure}

Let us consider the example in Fig. \ref{fig2}, in which the number of depots $m=3$, the number of drones is upper bounded by $n=3$, the barrier length $L=156$, the maximum drone path length $q=140$, and the depot coordinates $d_1=(18,10)$, $d_2=(78,10\sqrt{35})$ ($|dd_2|=|d_2h|=60$), $d_3=(138,10)$.

Since there are only 3 depots, the barrier will be divided into 3 or fewer parts, each of which is covered by drones from one depot. Since the optimal coverage preserves order, the drones of depot 1 should cover the left segment $[0,x]$, and the drones of depot 3 should cover the right segment $[y,L]$. Of course, in the optimal solution, it may turn out that some depots do not contain drones.

First, we find the minimum number of drones $n_i(a,b)\leq n=3$ in depot $i=1,2,3$ covering the segment $[a,b]$. Recall that one drone covers a segment of the barrier of length at least 1. The maximum segment that can be covered by drones from depot 1 is $[0,e]$, $e=78$. One drone from depot 1 covers the segment $[0,b]$ when $b\leq d=68$, but to cover the segment $[0,b]$, where $b\in[d+1,e]$, two drones from depot 1 are required. This means that $n_1(0,d)=1$, $n_1(0,b)=2$, $b=d+1,\ldots,e$, and $n_1(0,b)=+\infty$, when $b>e$. Next, we find $n_2(a,b)$ for $a\in[41,114]$ and all admissible $b>a$. In particular, $n_2(d,h)=1$. Since drones from depot 3 must cover the segment with the right boundary of $L$, we need the values $n_3(a,L)$. Depot $d_3$ is located symmetrically to depot $d_1$, therefore $n_3(a,L)=+\infty$ for $a<e$, $n_3(a,L)=2$ for $a\in[e+1,h-1]$, and $n_3(a,L)=1$ for $a\geq h$.

Let us find the minimum length of the paths $f_i(a,b)$ of drones from depot $i=1,2,3$ covering the segment $[a,b]$. For depot 1, since the covering preserves order, we are interested in the segments $[0,b]$. While $n_1(0,b)=1$, the length of the path of one drone $f_1(0,b)$ is equal to the perimeter of the triangle $0d_1b$. If $n_1(0,b)=2$, then 2 drones are used. In order for the total length of their paths to be minimal, they cover the segments $[0,c]$ ($d_1c$ is the perpendicular to $[0,L]$) and $[c,b]$, respectively, and $f_1(0,b)=|0d_1|+2|d_1c|+|d_1b|+|0b|$. For $b>d$ $f_1(0,b)=+\infty$. Next, we calculate $f_2(a,b)$ for admissible $a\in[41,114]$ and $b>a$. In particular, we have $f_2(d,h)=q=140$. And finally, we find $f_3(a,L)=|ad_3|+2|d_3g|+|d_3L|+|aL|$ for $a\in[e,h-1]$ and $f_3(a,L)=|hd_3|+2|d_3g|+|d_3L|+|aL|$ for $a\in[h,L-1]$.

Next, we calculate the values of the function $f(a,b)=\min_{i=1,2,3}f_i(a,b)$ for all integers $a\in[0,L-1]$ and $b\in[a+1,L]$. It is easy to understand that $f(a,b)=f_1(a,b)$ for $b\leq e$ and $f(a,b)=f_3(a,b)$ for $a\geq e$.

After applying algorithm $A_1$, we have a solution in which depots 1 and 3 have 2 drones each. The drones of depot 1 cover the segment $[0,e]$, and the drones of depot 2 cover the segment $[e,L]$, and the total length of the paths traveled by the drones is 340. But the number of used drones is greater than $n=3$. Therefore, this solution is infeasible and algorithm $A_2$ must be applied.

Algorithm $A_2$ considers the limitation on the number of drones. In our example, no depot can have more than one drone. Otherwise, the barrier will not be covered by three drones. Because of the symmetry of the outer depots, one drone from depot $d_1$ will cover the segment $[0,d]$, and one drone from depot $d_3$ will cover the segment $[h,L]$. The drone from depot 2 will cover the segment $[d,h]$. The total length of the drones’ paths is 420.

\section{Conclusion}
This paper considers the problem of monitoring (covering) a barrier (a segment $[0,L]$) on a plane. The coverage is performed by similar mobile devices (we call them the drones) with a limited energy reserve, which determines the maximum length of the drone’s path $q$. Each drone departs from its depot, moves along a segment of the barrier of length not less than 1 and returns to its depot, traveling a path no longer than $q$. The part of the barrier along which the drone moved is \emph{covered} by this drone. In the MinSum problem, it is required to place at most $n$ drones in $m$ depots, the coordinates of which are given, and to determine the trajectory of each drone in such a way that the barrier is covered, and the total length of the drones’ paths is minimal.

Previously, \cite{ErPlot20} considered the MinSum problem, when each depot contains one drone and it was necessary to determine the trajectories of the drones. For this version of the problem, there is not always an optimal coverage that preserves the order, so the dynamic programming algorithm does not always give an optimal solution. If the number of drones in each depot is the desired value (variable), then there is always an optimal coverage that preserves the order. In \cite{ErShad24}, a dynamic programming algorithm is proposed that solves the problem with complexity $O(mL^3)$, when the total number of drones is unlimited, and the boundaries of the segment covered by the drones of one depot are integers.

In this paper, we propose two algorithms, $A_1$ and $A_2$. Algorithm $A_1$ finds the optimal coverage by the \emph{optimal} number of drones $n^*$ with time complexity $O(L^2)$. If $n^*\leq n$, then this is the optimal solution to the MinSum problem. If $n^* > n$, then an $O(nmL^2)$-time algorithm $A_2$ is used, which constructs a feasible solution. Since each drone covers a segment of length at least 1, then $n\leq L$. Therefore, this complexity does not exceed the complexity of the algorithm proposed in \cite{ErShad24}, although it solves a more general problem.

To implement the proposed algorithms, the input functions are $n_i(a,b)$ -- the minimum number of drones in depot $i$ that cover the segment $[a,b]$, and $f_i(a,b)$ -- the minimum length of drone paths from depot $i$ that cover the segment $[a,b]$. Then, for algorithm $A_1$, the function $f(a,b)=\min\limits_{i=1,\ldots,m} f_i(a,b)$ was determined. Then the input length for algorithm $A_1$ is equal to $O(L^2)$ and the algorithm itself has linear complexity of the input length. However, calculating the values of functions $n_i(a,b)$ and $f_i(a,b)$, i.e. preprocessing, is quite time-consuming. In \cite{ErShad24} the functions $n_i(a,b)$ were not required, and all $O(mL^2)$ values of the functions $f_i(a,b)$ were found with time complexity $O(mL^3)$. In this paper, we reduce the complexity of calculating all values of the functions $n_i(a,b)$ and $f_i(a,b)$ to $O(nmL)\leq O(mL^2)$.

In this paper, as in \cite{ErShad24}, to determine the complexity we considered the case when drones cover segments $[a,b]$ with integer $a$ and $b$, and one drone covers a segment of length at least 1. Of course, the length of the segment covered by one drone can be at least some $\delta>0$. If $\delta$ is rational, then multiplying it by the corresponding integer leads us to the desired conditions. However, in the optimal solution, drones of one depot can cover a segment with non-integer end-points, and then the proposed algorithms construct approximate solutions. In the future, we plan to evaluate the accuracy of the solution constructed using the $\delta$ step and propose FPTAS.

\section*{Declarations}
\begin{itemize}
\item The manuscript was written on the initiative of the authors without financial support.
\item There are no conflicts of interest.
\item Ethical approval and consent for participation of the authors is available.
\item The authors agree to the publication of the manuscript.
\item All necessary data are in the text of the manuscript.
\item No additional materials are required.
\item The manuscript does not contain codes.
\item The first author is responsible for the formulation of the problem and the main ideas for the development and analysis of algorithms. The second author is responsible for the proof of some statements and the technical work.
\end{itemize}

\end{document}